# Artificial Intelligence in Experimental Approaches: Growth Hacking, Lean Startup, Design Thinking, and Agile

Parisa Omidmand[1], Saeid Ataei[2]

[1]Department of Economics, Texas Tech University

[2]Department of Systems Engineering, Stevens Institute of Technology

**Abstract**

Organizations increasingly adopt AI technologies to accelerate their performance and capacity to adapt to market dynamics. This study examines how organizations implement AI in experimental methodologies such as growth hacking, lean startup, design thinking, and agile methodology to enhance efficiency and effectiveness. We performed a systematic literature review following the PRISMA 2020 framework, analyzing 37 articles from Web of Science (WOS) and Scopus databases published between 2018 and 2024 to assess AI integration with experimental approaches. Our findings indicate that AI plays a pivotal role in enhancing these methodologies by offering advanced tools for data analysis, real-time feedback, automation, and process optimization. For instance, AI-driven analytics improves decision-making in growth hacking, streamlines iterative cycles in lean startups, enhances creativity in design thinking, and optimizes task prioritization in agile methodology. Furthermore, we identified several real-world cases that successfully utilized AI in experimental strategies and improved their performance across various industries. However, despite the clear advantages of AI integration, organizations face barriers such as skill gaps, ethical concerns, and data governance issues. Addressing these challenges requires a strategic approach to AI adoption, including workforce training, strict data management, and following ethical standards.

# Introduction

In recent years, organizations have turned to AI to increase business value through sustainable competitive advantage (Zebec & Stemberger, 2024). AI has a history of half a century; however, it gained momentum through rapid advances in computational capability and new technologies, including machine learning, deep learning, natural language processing (NLP), and computer vision, in recent decades (Mariani et al., 2023). AI has developed to the point that it enables automation and augmentation for decision-making and business process transformation. By accelerating overall organizational performance, AI has become a transformative force for organizations in various industries (Zebec & Stemberger, 2024). AI adoption generates opportunities for the company's fundamental shift in business operations and value creation approach.  AI technologies enable companies to manage uncertainty and increase cost efficiency and revenue (Åström et al., 2022). Human intelligence simulation of AI provides more informed decision-making, predictive analysis, and process optimization across various sectors. AI radically changes how businesses operate to deliver, create, and capture value (Mariani et al., 2023).

This study examines different AI technologies that support experimental approaches and enhance organizational performance. These approaches, including growth hacking, lean startup, design thinking, and agile methodology, rely on rapid iteration, innovation, and customer feedback analysis and can benefit significantly from AI's ability to analyze large datasets, automate repetitive tasks, and provide real-time insights (Soltan et al., 2025). Although prior research has examined AI's role in business process automation, decision-making, and innovation, the literature still lacks a focused synthesis of how AI can be leveraged across experimental approaches. This research provides a comprehensive analysis of how AI technologies integrate into these strategies while addressing the key challenges of AI adoption and proposing practical solutions for overcoming them. The primary questions of this research are which AI technologies provide benefits across these approaches, how AI can be leveraged to optimize them, and what risks and ethical considerations accompany their use.

# Background

## Growth Hacking

The growth hacking concept, introduced by Sean Ellis, combines marketing, data analytics, and product development to gain rapid growth in startups and established companies. This method contains continuous experimentation across the customer journey to identify the most effective ways to attract and retain customers. In this method, companies should transform data into actionable learning for continuous improvement to scale the business efficiently and effectively by acquiring and retaining active users (Bargoni et al., 2024). Growth hacking optimizes marketing activities using virtual marketing techniques, leveraging social media platforms, and implementing customer relationship management (CRM) tools. This method focuses on finding "hacks" or shortcuts that enable companies to grow significantly with minimum resources. Companies should implement resource-light and cost-effective tactics to target the right customers at the right time, at the right place, and in the most efficient ways.  This strategy is most effective in a digital landscape where companies can refine their strategies based on real-time data (Troisi et al., 2020).

Traditional growth hacking method relies on digital marketing channels such as social media, email marketing, and search engine optimization (SEO) for growth. These marketing tools are used in different stages of the "funnel" framework, including acquisition, activation, retention, revenue, and referral. This framework is central to growth hacking, enables customer journey optimization, and ensures they become loyal customers (Bargoni et al., 2024). There are several challenges associated with growth hacking, such as managing the complexity of data and ensuring the relevance of marketing strategy to rapid market changes. Moreover, this strategy's reliance on digital channels and platforms requires investment in digital marketing, e-commerce, and customer relationship management systems to leverage synergy and stay ahead of the latest trends and technologies. Growth hacking's interactive nature requires constant monitoring and adjustment through gathering real-time feedback and adapting quickly to market change for maximum impact (Petersen, 2024).

**Lean Startup**

The lean startup methodology is designed to allow entrepreneurs to develop and test their business ideas more quickly while lowering failure risk by emphasizing validated learning. This strategy focuses on creating a minimum viable product (MVP), which is a simplified version of the product that contains only essential features required to test the business hypothesis. This approach allows businesses, especially startups, to gain insightful feedback and rapid iteration based on data acquired through early adopter feedback (Raneri et al., 2023). The Build-Measure-Learn loop is the fundamental principle of lean startup, which guides the iterative product development process. This cycle includes building the MVP, measuring its performance through customer feedback and data analysis, and learning from results to make informed decisions about the next steps. By emphasizing validated learning, lean startup helps entrepreneurs speed up their development, test their business ideas, and reduce their failure risk. In sum, this methodology aims to achieve product-market fit as quickly as possible while minimizing waste (Harms & Schwery, 2020).

Lean startup methodology relies on market research, customer interviews, and testing to validate business ideas. This method helps startups to gain practical information to refine their products and business models. However, scaling this method raises challenges, particularly in large organizations where lean practices may conflict with existing structures and processes. The emphasis on speed is another difficulty that leads to a lack of depth in customer insights and results in products that do not fully meet market needs. Quality, quantity, and diversity of knowledge pool companies use affect their value creation process (Naeem et al., 2024). Additionally, the iterative nature of lean startup can pressure teams to deliver results quickly, which may lead to burnout and a lack of long-term strategic thinking. Knowledge from various and diverse knowledge bases allows innovators to increase the quantity, quality, and diversity of ideas and to create more value in their innovation processes—at a very low cost (Bouschery et al., 2023).

**Design Thinking**

Design thinking is an innovative approach focusing on recognizing customers' needs and desires. It is structured around five stages: empathize, define, ideate, prototype, and test. This method addresses end-user needs by developing solutions using creativity and problem-solving skills (Saeidnia & Ausloos, 2024). In the empathize stage, organizations acquire a deep understanding of users and their needs through interviews and observation. In the Define stage, they combine information gained in the empathize stage to identify the core problem that needs to be resolved. The Ideation stage is the process of generating creative ideas to address the defined problem. The next stage is prototyping, where ideas turn into tangible solutions. The ultimate stage is Testing, which involves getting user feedback on the prototypes and making adjustments based on them (Cautela et al., 2019).

Design thinking methodology involves extensive research, brainstorming sessions, and prototyping to ensure creativity, collaboration, and product and service development effectiveness. However, there is a challenge of managing the time and resources required through user research and balancing creative aspects with practical business constraints (Sreenivasan & Suresh, 2024). Businesses should ensure that prototypes have the potential to become scalable products. Although design thinking is an effective method to generate and test ideas, there is the challenge of a gap between conceptual products and the final product released to the market. The gap is difficult to bridge, especially when new designs should be implemented into an existing system and process. Moreover, the nature of design thinking can result in an extended development timeline, which may conflict with the need for speed in competitive markets (Cautela et al., 2019).

## Agile Methodologies

Agile methodology enables teams to adapt to change rapidly by prioritizing flexibility, collaboration, and continuous feedback. In a broader organizational sense, agility can also refer to the capability to respond quickly to unpredictable changes, which depends on effective coordination and information integration across actors (Dubey et al., 2022). In this study, however, agile is primarily discussed as a project and innovation methodology used to improve responsiveness, iteration, and team performance.

Traditionally, agile methodology was used in software development to manage complex projects by breaking them into smaller manageable tasks. The agile methodology then spread to various industries where they faced the necessity of flexibility and responsiveness to customer needs (Ameen et al., 2024). Companies need to adopt their strategy while maintaining this balance between speed and quality to create cost-effective, high-quality, and on-demand products and services and stay competitive by responding to customer demand (Mrugalska & Ahmed, 2021). The agile methodology is based on teamwork and requires proper task assignment, team orientation, commitments, training, and team structure which cannot be achieved without proper management. Moreover, the practice of the agile method in big organizations is more challenging to provide effective communication and collaboration between teams and team members (Song et al., 2022). Having outlined the fundamental principles and challenges of these four experimental approaches, this study now examines how AI technologies are being integrated into each methodology through a systematic review of academic literature.

## Methodology

This literature review explores AI's application in experimental approaches such as growth hacking, lean startup, design thinking, and agile methodology. A systematic review was conducted using the PRISMA 2020 framework, which is designed to provide a comprehensive and transparent process of identification and selection of relevant studies. This review includes articles and review articles published between 2018 and 2024, focusing on AI advancement and its integration into organizations and experimental methods. Additional sources published in 2025 were used only as supporting literature to contextualize the discussion and were not part of the PRISMA-reviewed sample. The review aims to answer three primary research questions: (1) Which AI technologies provide benefits across these experimental methodologies? (2) How can AI be leveraged to optimize experimental approaches (growth hacking, lean startup, design thinking, agile) for maximum innovation and efficiency? (3) What are the potential risks and ethical considerations of using AI across these methodologies?

The literature was sourced from two major academic databases: Web of Science (WOS) and Scopus. The search began with combinations of keywords in article titles, which yielded 1,424 records from WOS and 2,042 from Scopus. These combinations were "AI and Growth Hacking," "AI and Design Thinking," "AI and Agile," "AI and lean startup," "AI and Innovation," and "AI and Business Model Innovation." After that, we set some inclusion and exclusion criteria to restrict the number of articles and get the most relevant document for the final research.

The review applied strict inclusion and exclusion criteria. The included articles were published between 2018 and 2024, focused on AI implications in the business or management sector, and discussed AI adoption in experimental methods such as agile, growth hacking, lean startup, and design thinking. The excluded articles were irrelevant to AI application or focused on its application in fields such as finance, human resource management, leadership, environment, and sustainability, with a lack of connection to AI in the experimental approach. After eliminating articles published before 2018 and those related to other sectors, we ended up with 243 sources from WOS and 205 from Scopus. In the next step, we removed 133 duplicate articles and ended up with 315 articles for screening.  After carefully screening the 315 articles, we excluded 187 records due to lack of relevance to the main topic, leaving 128 reports sought for retrieval. Of these, 84 reports were not retrieved. The remaining 44 reports were assessed for eligibility, and 7 were excluded after full-text review, resulting in a final sample of 37 studies.

The collected literature was systematically reviewed and synthesized to identify themes and patterns related to AI's application in experimental approaches. The final selection of 37 studies was analyzed to identify AI technologies that strengthen experimental methods and enhance business production, marketing, innovation, and overall performance. The key findings highlight AI-driven opportunities that improve the effectiveness of experimental approaches, identify barriers related to organizational knowledge gaps, privacy, and ethics, and propose ways to integrate AI more responsibly into organizational practice.

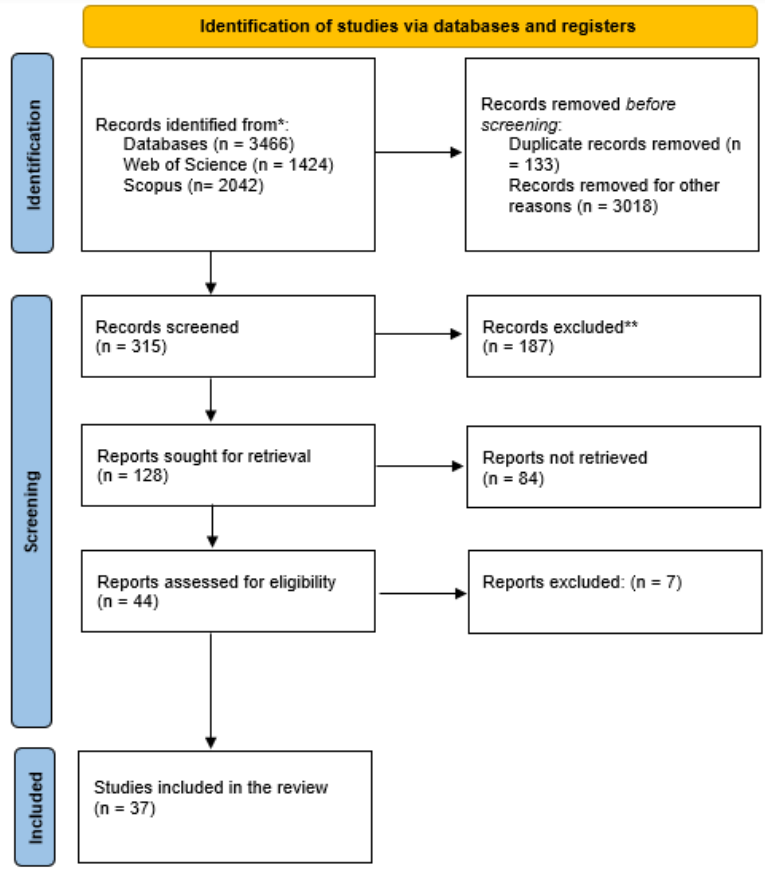

**Figure 1. PRISMA 2020 flow diagram**

## Findings

### AI Integration in Experimental Approaches

### AI in Growth Hacking
Growth hacking methodology has been revolutionized by AI-driven analytics and data mining, which enabled organizations to process vast amounts of data and extract valuable insights from them (Omidmand, 2025). Combining the strength of big data and continuous learning from experiments enables companies to adapt to an ever-shifting competitive environment (Bargoni et al., 2024). Digital technologies such as digital channels and platforms have a central role in reaching global customers effectively. Implementing digital marketing, e-commerce, and customer management systems helps organizations leverage growth hacking synergies and increase customer acquisition, engagement, and relationships in international markets (Petersen, 2024). Techniques such as natural language processing and machine learning enable the analysis of unstructured data from different sources and provide valuable insights into customer preferences. They also predict customer preferences through algorithms and evaluate the performance of existing channels to identify the most effective one (Santoro et al., 2024). Utilizing NLP and ML in customer behavior prediction and channel optimization results in effective digital marketing strategies (Jin et al., 2024).

AI enables personalized marketing based on customer behavior, preferences, and demographics. For instance, big data tools retrieve, process, analyze, and report customer online opinions and behaviors to help companies create tailored products and services and also improve their business model innovation (Mariani et al., 2023). AI algorithms exploit information exchange between businesses and customers, resulting in a co-specification of products or services. These algorithms rely on broad numeric and non-numeric databases for given customers, which can be compared with other customers' data and provide high-level, objectively tailored results where customers' preferences constantly evolve (Grandinetti, 2020).

AI systems empower growth hacking by facilitating user acquisition and retention through customization. Tools like cognitive computing systems (machine learning, decision support systems, and group decision-making), API (Application Programming Interface), and web crawlers are trained to understand technical and industry-specific content. They use advanced reasoning and predictive modeling to create and deliver personalized experiences to customers (Troisi et al., 2020). Organizations increase their conversion rate and customer satisfaction using AI-driven data analysis of user interaction, product view, and purchase history to tailor personalized offers and recommend them to users (Santoro et al., 2024). AI systems analyze historical data, market trends, and external factors to provide accurate predictions, optimizing inventory, server capacity, and marketing efforts. AI-automated analyses of customer behavior enable targeted product and service suggestions and campaigns designed based on customer desires and enhance product-market fit. This analysis facilitates strategy adjustments and leads to effective communication campaigns, re-targeting, and grants a higher engagement rate (Santoro et al., 2024).

IKEA efficiently integrated AI into its growth hacking strategy. This company trained sales, distribution, and pricing employees to utilize new technologies, analyze data, and leverage AI to enhance the online shopping experience and customer satisfaction. IKEA streamlined its operations with cloud migration and optimized its e-commerce website with AI-enabled systems to optimize product recommendations and customer services and provide a unique online shopping journey (Esmaeili et al., 2024). Moreover, IKEA developed an augmented reality (AR)- powered application called Place app, which allows customers to try the products and experience the purchasing process virtually. This integration enhanced company customer acquisition and retention while helping customer data collection for further optimization (Bargoni et al., 2024).

## AI in Lean Startup

Research-driven online review platforms (RORP) generate novel insights through a combination of science and large scale, low cost, fast speed, and complex digital experiments using real-world customers on platforms with AI-based big data analysis capabilities. RORP emphasizes experimentation and discovery, allowing radical changes in business models (Mariani & Nambisan, 2021).

AI-powered analysis tools allow organizations to gather customer feedback in real time and refine their response to customer needs (Cautela et al., 2019). AI efficiently analyzes feedback through different channels, such as email surveys, and provides qualitative and quantitative insights (Santoro et al., 2024). Moreover, Cognitive Insight (CI) increases context awareness, learning, and analysis and enables organizations to detect and interpret patterns in data (Zebec & Stemberger, 2024). AI-driven analysis of customer feedback helps managers shape product and service strategies to align with market acceptance and adaptation rates (Naeem et al., 2024). Several artificial intelligence techniques, such as feature extraction, sentiment analysis, anomaly or novelty detection, and time series analysis, enable effective customer review analysis and product feature classification. AI algorithms leverage user rating analysis of new products, predicting user opinions and detecting design decisions that lead to convincing results. This AI-driven analysis provides an opportunity to learn for the companies and results in effective and efficient application of the BML cycle (Raneri et al., 2023; Omidmand et al., 2025).

Language models can draw on a wide range of topics that may fall outside team members' immediate expertise and can therefore expand innovation opportunities. They can generate useful outputs quickly, although the quality of those outputs still depends on task framing, human judgment, and contextual interpretation (Maleki et al., 2024). Transformer-based language models further enable the integration of diverse knowledge bases and can improve the quality and quantity of ideas generated in innovation processes at relatively low cost. Human actors can integrate AI into their process as new colleagues by providing language models (LMs) with their knowledge and ideas through natural language. AI's ability to connect concepts and ideas plus human knowledge leads to improvement in new product development (NPD) practices (Bouschery et al., 2023). An increasing number of organizations rely on predictive analysis generated by the customer-centered research-driven online review platform (RORP) to test new concepts, products, pricing, promotions, and advertisements. Virtual and Augmented Reality (VAR) also help RORP to conduct online experiments and test customer acceptance of products. As a result, accurate analysis and insight are generated through the customer selection of options and identifying the right product or service, the right price, and the segment to target (Mariani & Nambisan, 2021).

## AI in Design Thinking

AI influences design thinking by supporting decision-making, prototyping, and ideating and testing processes, leading to creative and effective design solutions and eliminating dull processes. Design thinking characteristic of coming up with fresh and innovative ideas by challenging presumptions has harmony with AI's ability to process massive amounts of data and pattern identification to provide creative solutions to problems (Sreenivasan & Suresh, 2024). Moreover, AI analyzes user behavior drawn from data sources such as social media, websites, and mobile apps and offers scientific and objective qualitative input to the design process. AI-driven team building and task management reinforce the collaborative and collective view of creativity and guarantee a holistic view of designing problems (Cautela et al., 2019). Successful innovation needs the availability of a high volume of good-quality data, which AI technology can provide nowadays. Organizations use AI for AI-driven innovation in different ways including AI-driven platforms to explore innovative solutions, AI-powered idea generation tools that use customers' data and trends to design innovative concepts, and AI algorithms that enable brainstorming and refining creative ideas through automated suggestion systems (Prem, 2019; Saeidnia & Ausloos, 2024). By eliminating typical scale, scope, and learning barriers of human-intensive design, AI offers better performance of customer-centric innovation rate (Verganti et al., 2020). There are different AI tools that accelerate the innovative potentials of an organization, such as AI pattern discovery which increases understanding of the problems, image processing apps that analyze competitor products, and language processing software that empower user observation and analysis capability of the creative team to match them with different users (Cautela et al., 2019).

Integrating AI with Design thinking allows focused data-driven insights with empathetic understanding and enhances its human-centered approach. Evaluating user data and feedback at scale helps designers find the trends and preferences guiding their work. This approach guarantees that the solution illustrates users' motivation, goals, and demands (Saeidnia & Ausloos, 2024). AI's ability to target different customer segments and locations leads to openness in solutions, reduces complexity in developments, and brings functionality to address the needs of various operating markets. AI-based recommendation engines can use these customers' data to generate real-time personalized offers automatically (Hasan & Ojala, 2024). AI-based quantitative analysis, combined with design thinking's qualitative insights, helps develop user-centric solutions and create significant value for end users (Sreenivasan & Suresh, 2024). Users' emotions have always been key in the design thinking process for generating new solutions. AI allows access to statistical evidence on user emotions by investigating behaviors in context and recreating user behavior for analysis, eliminating creative teams' biases. Traditionally, companies gained emotional dimension in design through interviews or direct observation, which includes bias from subjective understanding of the observer. However, the AI system's access to a high volume of less biased data on user behavior, emotion, and psychological responses modifies the empathic dimension of design thinking (Cautela et al., 2019).

AI speeds up the interpretation and integration of data from different sources and automates a large part of the prototyping and learning process (Sreenivasan & Suresh, 2024). Web apps and platforms utilize AI-based solutions that automatically run user tests and visual corrections. Moreover, AI simulates user behavior and implements virtual prototype sessions that automatically provide advanced solutions while reducing some forms of human bias. This shift in using AI leads prototyping to the phase where products and services are tested not by humans but by robots or intelligent agents. Designers test solutions on virtual individuals characterized by real emotions and preferences, which gives them more time and energy to focus on ideation (Cautela et al., 2019). Prototyping phases can get AI tools into practice to generate mockups and interactive prototypes. These tools can also optimize these prototypes based on user interaction and predictive modeling and implement rapid testing to gather feedback on their efficiency. AI-driven user testing tools can evaluate user experiences and gather insights to refine design solutions. In addition, AI algorithms enable testing designs and comparing variations to find the most effective one. The fusion of AI with the prototyping and testing process offers new possibilities for innovation (Saeidnia & Ausloos, 2024).

Integrating AI technology into the design thinking approach is a key point in Airbnb's rapid and successful response to user needs and maximized profit. AI-driven analysis of user interactions, including time spent on listing pages and behaviors during the booking process, allows capturing how guests interact with this platform, what should be focused on, and generating solutions for each user to refine their next search experience. Airbnb uses AI to optimize interactions with guests and hosts as a two-sided platform. For hosts, AI dynamically adjusts prices based on factors like booking history, listing popularity, and check-in trends, ensuring real-time adjustments to meet market demand. This dual problem-solving loop enables personalized solutions, enhancing user satisfaction and adapting to their needs. AI-driven design thinking has helped Airbnb empathize with users, identify challenges, provide tailored solutions, improve service quality, and achieve personalization (Verganti et al., 2020).

## AI in Agile Methodologies

Artificial intelligence provides powerful tools to address limitations and enhance agility in project management. Integrating AI into agile workflows enhances project management capabilities by automating repetitive tasks, optimizing resource management, predicting and mitigating risk, improving estimation accuracy, and providing real-time project insight (Ameen et al., 2024). For instance, generative AI redefines the skills required for white-collar tasks and reshapes workers' roles from creator to editor (Kanbach et al., 2024). This integration enhances the efficiency and responsiveness of project teams and allows them to concentrate on high-value activities and make informed decisions (Gama & Magistretti, 2025).

Organizational agility improvement requires several ongoing practices including communication, collaboration, engagement, and organizational support (Shafiabady et al., 2023). AI allows the quick collection and process of complex information, which is necessary to improve performance and information integration among humanitarian actors (Dubey et al., 2022). Moreover, AI enables diversity, cross-functional, and interdisciplinary collaboration through diverse talent recruitment and team formation, and it enables organizations to respond quickly by addressing challenges to individuals and teams capable of solving them (Yams et al., 2020). This integration increases transparency, information alignment, and real-time collaboration, leading to a more agile organization (Dubey et al., 2022). Organizations are using AI-assisted human teams as a strategy to strengthen AI and human integration. These teams consist of human and AI agents that boost agility by alerting team members to change, reducing uncertainties through constant learning, and supporting leadership that facilitates information exchange and decision-making. They also implement a flat team structure with open AI-powered communication channels, resulting in team structure changes, facilitating information flow, and improving agility in the organization (Song et al., 2022). AI technology capability for data creation and communication activities enhances the adaptability of organizations. For instance, AR and VR technologies significantly help organizations control activities by automating manufacturing and supply chain activities and improving stakeholder collaboration. Moreover, the connection of objects over internet platforms makes information systems help them quickly respond to changes. AI's ability to collect data and subsequently analyze them enhances organizations' self-learning process and helps them be flexible and agile (Mrugalska & Ahmed, 2021).

Companies all over the globe are leveraging AI capabilities to enhance their agility and responsiveness to dynamic market environments. For instance, Amazon increasingly utilizes AI-enabled analytic tools, platforms, and channels to monitor and communicate with its supply chain and acts rapidly in response to changes or disruptions. This organization understands the importance of agility in quick response to any market change and identifying which services and products match customer needs, increasing customer satisfaction and thus creating value for the company (Fosso Wamba, 2022).

## Integration Strategies

Several practices can guide the integration of AI into these approaches. For instance, starting with small, scalable projects allows the incremental adoption of AI in business practices with limited risk of large-scale disruption, such as automated customer feedback analysis or optimized resource management through predictive analysis (Gama & Magistretti, 2025). Another aspect to consider in successful integration is data quality and governance. As AI systems depend on accurate data to generate meaningful insight, organizations need to guarantee their data collection process is standard and carefully managed, which provides reliable outcomes (Soori et al., 2024). AI-driven optimization models can enhance operational decision-making across various business functions, from supplier selection to resource allocation, ensuring efficiency and cost-effectiveness (Shahrokhi et al., 2021).

In addition, training and upskilling the workforce is critical for the efficient integration of AI systems into organizations, enabling employees to use AI tools comfortably and understand how to leverage them effectively (Bencsik, 2021). Management's leadership views about AI would establish the culture for the rest of the team. Managers can effectively encourage employees to adopt AI-supportive behaviors and help establish a data-driven culture across different stages of the business process. These practices will ensure a conducive environment to facilitate adoption (Chatterjee et al., 2021). Companies should also establish clear guidelines for data management and analysis, ensuring that AI tools are used in a way that aligns with business goals and ethical standards. In this way, companies can build experience, trust, and confidence in AI systems before expanding AI implementation (Saeidnia & Ausloos, 2024).

## Challenges and Ethical Considerations

There are several technical and organizational challenges regarding AI adoption for experimental approaches. Integrating AI into existing organizational strategies and processes can be complex and often requires investments in infrastructure, tools, and workforce development (Song et al., 2022; Mirzaei et al., 2025). Additionally, AI models rely on a large amount of high-quality input data to provide effective output and solutions, which is challenging as many organizations struggle with fragmented or incomplete datasets (Bouschery et al., 2023). Another notable challenge is the skill gap between companies' internal expertise in AI system management and the required knowledge in data science, machine learning, and AI technologies. This is more obvious where real-time decision-making and fast adjustments are critical (Ameen et al., 2024).

While AI technology has great potential for business innovation, organizations must carefully consider the ethical implications of its application, including governance, transparency, accountability, and privacy (Rafel et al., 2023). AI-driven systems may reflect biases in their training data, making it essential to utilize fairness-aware algorithms, diverse data sets, and regular audits to ensure transparency. Organizations should also ensure data privacy and compliance with ethical and legal standards to mitigate these biases and integrate AI into businesses responsibly. Collaboration with experts in ethics, sociology, and human rights is crucial to identifying potential blind spots in AI applications and ensuring ethical and social values have been incorporated (Sreenivasan & Suresh, 2024). Moreover, implementing guidelines and safeguards in AI-related ethical concerns such as privacy, bias, transparency, and accountability must be prioritized to protect user rights and ensure unbiased outcomes (Saeidnia & Ausloos, 2024). AI systems require proactive measures throughout their life cycle to guarantee fairness, transparency, and accountability. The explainability in design (EID) methodology systematically helps the design team address explainability-related issues while encouraging critical thinking and reducing barriers to engaging in ethical AI decisions. Value Sensitive Design (VSD) also incorporates values like fairness and privacy into human-computer interaction workflows (Zhang & Yu, 2023).

## Limitations and Suggestions for Future Studies

Like all research, our systematic literature review displays some limitations. First, we collected data from Scopus and Web of Science (WOS) over other databases like Google Scholar. Future research may collect data from different databases, thus gathering further research outputs. Second, as a systematic review, our keywords and inclusion and exclusion criteria may have influenced the identification of relevant articles. Although we tried to be comprehensive, the literature considered is limited, and the inclusion of other keywords, categories, research type, and publication year could yield a more comprehensive result. Third, a more granular analysis of the sample articles could be carried out, resulting in a more comprehensive outcome. Finally, it is important to note that AI technology is evolving rapidly, and research on its application in organizational strategies continues to expand. Therefore, there is a need to continuously explore new AI methods and their application to experimental approaches.

## Conclusion

This systematic literature review of 37 studies demonstrates that AI integration fundamentally enhances experimental methodologies across multiple dimensions. Addressing our three research questions, this review identifies machine learning, natural language processing, computer vision, and transformer-based language models as the core AI technologies benefiting all these methodologies. These technologies optimize experimental approaches through automation of repetitive tasks, real-time data analysis, and personalization at scale. However, organizations face technical risks including data quality challenges and algorithm bias, organizational risks such as skill gaps, and ethical risks involving privacy and accountability concerns. In growth hacking, AI-driven analytics and personalization enable organizations to optimize customer acquisition and retention strategies. For lean startup methodology, AI accelerates the Build-Measure-Learn cycle through real-time feedback analysis and predictive modeling. In design thinking, AI augments creativity by processing vast datasets while reducing bias in user research. Finally, in agile methodologies, AI automates repetitive tasks and enhances team collaboration and responsiveness.

Although AI technology offers significant advantages, its integration into these approaches is not without challenges. The most important issues are data quality, ethical considerations surrounding bias and transparency, and organizational skill gaps. To address these challenges, organizations require proactive approaches, including improved data governance, continuous workforce training, and the development of ethical frameworks that ensure responsible AI use. Organizations must ensure that AI systems are aligned with business objectives and ethical standards to manage risk and avoid unintended consequences. However, organizations should balance the benefits of AI with its ethical and practical risks to demonstrate social and technological responsibility. As AI continues to evolve and expand its influence, businesses must adopt a well-designed strategic plan for AI integration, starting with small, scalable projects, ensuring robust data governance, and fostering a culture of continuous learning and ethical awareness to mitigate the risk of failure. Future research should explore new AI applications and their effect on organizational performance. Additionally, they should investigate successful AI adoption strategies while addressing its challenges. Ultimately, AI's integration into experimental approaches promises to redefine organizational processes and performance, positioning companies to succeed in a competitive and fast-paced digital landscape.

chain: A practice-based view. International Journal of Production Economics, 250, Article 108618. https://doi.org/10.1016/j.ijpe.2022.108618

9. Esmaeili, M., Ahmadi, M., Ismaeil, M. D., Mirzaei, S., & Canales Verdial, J. (2024). Advancements in AI-Driven Customer Service. In 2024 IEEE World AI IoT Congress (AIIoT). IEEE. https://doi.org/10.1109/AIIOT61789.2024.10579008

10. Fosso Wamba, S. (2022). Impact of artificial intelligence assimilation on firm performance: The mediating effects of organizational agility and customer agility. International Journal of Information Management, 67, 102544. https://doi.org/10.1016/j.ijinfomgt.2022.102544

11. Gama, F., & Magistretti, S. (2025). Artificial intelligence in innovation management: A review of innovation capabilities and a taxonomy of AI applications. Journal of Product Innovation Management, 42(1), 76–111. https://doi.org/10.1111/jpim.12698

12. Grandinetti, R. (2020). HOW ARTIFICIAL INTELLIGENCE CAN CHANGE THE CORE OF MARKETING THEORY. INNOVATIVE MARKETING, 16(2), 91–103. https://doi.org/10.21511/im.16(2).2020.08

13. Harms, R., & Schwery, M. (2020). Lean Startup: Operationalizing Lean Startup Capability and testing its performance implications. Journal of Small Business Management, 58(1), 200–223. https://doi.org/10.1080/00472778.2019.1659677

14. Hasan, R., & Ojala, A. (2024). Managing artificial intelligence in international business: Toward a research agenda on sustainable production and consumption. THUNDERBIRD INTERNATIONAL BUSINESS REVIEW, 66(2), 151–170. https://doi.org/10.1002/tie.22369

15. Jin, K., Zhong, Z., & Zhao, E. (2024). Sustainable Digital Marketing Under Big Data: An AI Random Forest Model Approach. IEEE TRANSACTIONS ON ENGINEERING MANAGEMENT, 71, 3566–3579. https://doi.org/10.1109/TEM.2023.3348991

16. Kanbach, D., Heiduk, L., Blueher, G., Schreiter, M., & Lahmann, A. (2024). The GenAI is out of the bottle: Generative artificial intelligence from a business model innovation perspective. REVIEW OF MANAGERIAL SCIENCE, 18(4), 1189–1220. https://doi.org/10.1007/s11846-023-00696-z

17. Maleki, E., Chen, L.-T., Vijayakumar, T. M., Asumah, H., Tretheway, P., Liu, L., Fu, Y., & Chu, P. (2024). AI-generated and YouTube videos on navigating the U.S. healthcare systems: Evaluation and reflection. International Journal of Technology in Teaching and Learning, 20(1), 40–72. https://doi.org/10.37120/ijttl.2024.20.1.03

18. Mariani, M., Machado, I., Magrelli, V., & Dwivedi, Y. (2023). Artificial intelligence in innovation research: A systematic review, conceptual framework, and future research directions. Technovation, 122, Article 102623. https://doi.org/10.1016/j.technovation.2022.102623